# On the lack of relation between physics and "Quantum discord and Maxwell's demons"

Elias P. Gyftopoulos
Massachusetts Institute of Technology
77 Massachusetts Avenue, Room 24-111
Cambridge, Massachusetts  02139 USA

The information-theoretic arguments presented in a recent publication on "Quantum discord and Maxwell's demons" are discussed, and found not to address the problem specified by Maxwell.  Two interrelated and definitive exorcisms of the demon, one purely thermodynamic, and the other quantum-thermodynamic are briefly discussed.  For each of the two exorcisms, the demon is shown to be incapable to accomplish his assignment neither because of limitations arising from information-theoretic tools at his disposal, nor because of the value of his IQ.  The limitations are due to the physics of the state of the system on which he is asked to perform his demonic acts.

PACS numbers: 05.70-a, 05.70.Ln, 03.65-w, 05.30-d

I.      INTRODUCTION

In "Quantum discord and Maxwell's demons" [1], the author introduces many ideas that may be useful in information theory but are relevant neither to physics (thermodynamics, and quantum-thermodynamics), nor to Maxwell's demon – there is only one and not several demons.  In what follows, I present specific theoretical and/or experimental objections to statements appearing in [1], and then summarize briefly two intimately interrelated, nonstatistical, rigorous, noncircular, and definitive exorcisms of the demon, one purely thermodynamic, and the other quantum-thermodynamic.  At the risk of being immediately characterized as ignorant and arrogant, I assert with one hundred percent confidence that the two definitive exorcisms are the only ones that address and respond to the problem posed by Maxwell, and they differ radically from every discussion of the demon that has appeared in books and the archival literature until 2002.

Objectionable statements are discussed in Section II, a thermodynamic definitive exorcism in Section III, a quantum-thermodynamic exorcism in Section IV, and concluding remarks in Section V.

II.     OBJECTIONABLE STATEMENTS

1.      Zurek [1] claims that: "Maxwell's demon was introduced to explore the role of information and, more generally, to investigate the place of intelligent observers in physics. …The role of the demon is to implement an appropriate conditional dynamics – to react to the state of the system as revealed through its correlation with the state of the apparatus".

These claims represent neither the nature of the demon, nor the tasks and results he is asked to accomplish.  Specifically, all physical theories are based on information or, more precisely, what Margenau [2] calls perceptions.  Such perceptions are assumed to be objective and not information-theoretic.

In the context of classical mechanics, the only theory that was known to Maxwell during his lifetime, individual particles do not have an intrinsic thermodynamic property called entropy.  Accordingly, an omniscient and omnipotent observer – a demon – can make observations at the atomic or molecular level that are not delimited by the laws of thermodynamics.  Moreover, the regularization of observations made by such a demon can be accomplished only by the best and the brightest amongst us, such as Newton, Maxwell, Einstein, Planck, and Heisenberg. So the demon was not introduced "to explore the role of information, and the place of intelligent observers in physics". He is representative of the scientific method employed by all individuals deserving the title of scientist.

2.　　Zurek [1] claims that: "a quantum demon – an entity that can measure nonlocal states, and implement quantum conditional operations – could be more efficient than a classical one. I show that quantum demons can extract more work than classical demons from correlations between quantum systems, and that the difference is given by the *quantum discord*, ... a measure of the quantumness of correlations".

These claims are faulty for at least three reasons. First, there exists no such a concept as a nonlocal state. The term state is defined for a system, and a system does not have local parts. The reason for this objection is discussed in the next subsection. Second, the realm of classical mechanics does not overlap with that of quantum theory. The former is a limiting case of the latter at relatively high values of energy [3]. Said differently, the effects of entropy, the most important distinguishing feature between mechanics and thermodynamics, are negligible in the realm of classical mechanics [4]. In contrast to the special theory of relativity which becomes numerically important for very high energies, the new thermodynamics is numerically important for relatively low energies because then and only then the effects of entropy on available energy (exergy) are numerically sizeable. The reason is that at high energies and thermodynamic equilibrium, energy is proportional to temperature, whereas entropy is proportional to the logarithm of temperature.

The third reason is the extraction of work. If such extraction were possible even temporarily, then Maxwell's expectations would be verified, and the laws of thermodynamics would be violated at the molecular level.

3.　　Zurek [1] introduces the concept of "quantum discord". This concept violates the complete, noncircular, and universal definitions of the terms system, property (at an instant in time), and state.

The definition of system requires that the entity qualified as a system be separable from and uncorrelated with other systems in its environment. This definition does not apply to systems $A$ and $S$ shown in Figure 1 of Ref. [1], and therefore one cannot define properties of $A$ and $S$ in the presence of correlations. Said differently, $A$ plus $S$ are one system $C$, and all descriptions and conclusions must refer to $C$, and only $C$.

Incidentally, neglect of correlations, or techniques to get around them can lead to strange conclusions. An outstanding, and easily understandable example is the collision integral used in the apparent *proof* of Boltzmann's H-theorem. The neglect of correlations results in an apparent increase of entropy derived from classical mechanics, that is, from a theory that applies only to zero-entropy physics.

4.　　Zurek [1] claims that: "in quantum theory,…a measurement will, in general, redefine the state of the measured object,…, and that this results in an increase of the entropy that an observer attributes to a pair of correlated systems".

It is true that under the assumptions made by Zurek, the entropy of a pair of systems increases. The assumptions are based on von Neumann's collapse of a wave function or, equivalently, what is called the projection postulate. However, as a result of an excellent, rigorous and complete analysis, the projection postulate is shown to be at once "absurd, false, and useless" [5].

The only thing that I wish to add here is an argument against the postulate based on a violation of the position-momentum uncertainty relation. Consider a structureless particle confined in a one-dimensional, infinitely deep potential well of width $L$. Initially, the particle is in a state characterized by a projector $|\psi\rangle\langle\psi|$. According to the projection postulate, upon a momentum measurement the particle must collapse into a momentum eigenstate. Suppose that the eigenstate just cited is characterized by the ith momentum projector $|p_i\rangle\langle p_i|$. For such a projector, the standard deviation of position measurement results satisfies the relations $0 < \Delta x < L$, and the standard deviation of momentum measurement results $\Delta p = 0$. Accordingly, $\Delta x \Delta p = 0 < \hbar/2$ instead of $\Delta x \Delta p \geq \hbar/2$. In view of the unquestionable validity of uncertainty relations, we must conclude that the projection postulate cannot be a valid postulate of quantum theory.

5. Zurek [1] claims that: "because of the cost of erasure…neither classical nor quantum demons can violate the second law". This claim misrepresents Maxwell's specifications, and has no relation to physics. Maxwell's specification of his demon is that of an entity that can do whatever he pleases without any expenditure, including any work–energy only– expenditure, because in classical mechanics no expenditure is required in the course of measurements. Accordingly, if someone conceives of a method that tries to achieve what the demon is supposed to do but at some cost–any kind of cost–then that method is not a Maxwellian demon. Incidentally, all the methods that have appeared in the archival literature until December 2002 are "costly", and, therefore, do not address the problem posed by Maxwell.

It is interesting and informative to compare Maxwell's expectations of his demon with Sadi Carnot's seminal ideas that resulted in the discovery of the maximum thermal efficiency of an engine operating between two reservoirs at temperatures $T_1$ and $T_2$, respectively, or equivalently 100% thermodynamic efficiency or effectiveness. He said [6]: "The question has often been raised whether the motive power of heat is unbounded, whether the possible improvements in steam engines have an assignable limit–a limit by which the nature of things will not allow to be passed by any means whatsoever, or whether, on the contrary, these improvements may be carried on indefinitely. …In order to consider in the most general way the principle of the production of motion by heat, it must be considered independently of any mechanism or any particular agent. It is necessary to establish principles applicable not only to steam engines but to all imaginable heat engines, whatever the working substance, and whatever the method by which it is operated. … Machines which do not receive their motion from heat, those which have for a motor the force of men or of animals, a waterfall, an air current, etc., can be studied even to their smallest details by the mechanical theory. All cases are forseen, all imaginable movements are referred to these general principles, firmly established, and applicable under all circumstances. This is the character of a complete theory. A similar theory is evidently needed for heat engines. We shall have it only when the physics shall be extended enough, generalized enough, to make known beforehand all the effects of heat acting in a determined manner on any body."

III.     A THERMODYNAMIC DEFINITIVE EXORCISM

The thermodynamic definitive exorcism of the demon [7] is based on an exposition of thermodynamics in which entropy is proven to be an intrinsic, nondestructible, nonstatistical property of any system (both large and small), in any state (both thermodynamic equilibrium and not thermodynamic equilibrium), in the same sense that inertial mass is an intrinsic property of any system in any state[8]. In this exposition, it is shown that the demon cannot accomplish his task because he is asked either to reduce the entropy of a given amount of air without compensation or, equivalently, to extract work (only energy) from air under conditions that require the simultaneous extraction of both energy and entropy. So, the insurmountable limitations on his actions are dictated neither by the procedures and equipment at his disposal nor by the effects of irreversibility. They result from the characteristics of the properties of the air on which he is asked to perform his demonic acts.

Maxwell did not consider the characteristics just cited because his universal paradigm of physics was classical mechanics (the word paradigm is used with the meaning coined by Kuhn [9]). Accordingly, he interpreted correctly statistics as necessary tools to bypass difficult dynamic calculations, and mistakenly restricted the use of these tools to macroscopic systems. Thus, a limitation was created about the validity of one of the then prevailing statements of the second law of thermodynamics. In the new exposition, the limitation is eliminated because the laws of thermodynamics are part of a universal paradigm, classical mechanics is a special case, and the entropy theorem is nonstatistical and valid at all levels, both macroscopic and microscopic.

IV.    A QUANTUM-THEORETIC DEFINITIVE EXORCISM

The quantum-theoretic exorcism of the demon [11] is based on a unified quantum theory of mechanics and thermodynamics in which: (a) the probabilities represented by a density operator $\rho$ $(\rho \geq \rho^2)$ are exclusively quantal and not a mixture of quantal probabilities derived from projectors $\rho_i$ $(\rho_i = \rho_i^2)$, and statistical (informational) probabilities $\alpha_i$ expressing lack of knowledge of the state of the system; said differently, $\rho$ is represented by a homogeneous ensemble of identical systems, identically prepared; *homogeneous* is an ensemble in which every member is assigned the same density operator $\rho$ as any other member, that is, experimentally (as opposed to algebraically) the ensemble cannot be decomposed into statistical mixtures of either projectors or other mixtures; (b) the value $\langle A \rangle$ of any observable is not any particular measurement result–any eigenvalue of the Hermitian operator A representing the observable–but the average of an ensemble of measurement results, that is $\langle A \rangle = \Sigma_i \, a_i / N$, where $a_i$ is the measurement result of A from the ith member of the ensemble, and $N \to \infty$, that is, the number of members of the ensemble; and (c) in a thermodynamic equilibrium state, it is shown that each molecule has a zero value of momentum, that is, each molecule does not move.

It is noteworthy that, in principle, expectation values are measureable, and a complete set of expectation values of linearly independent observables represents a linear mapping of a density operator that characterizes a homogeneous ensemble. Conversely, a density operator that characterizes a homogeneous ensemble is a linear mapping of a complete set of linearly independent expectation values.

It follows that, in the context of the unified quantum theory of mechanics and thermodynamics, the demon is shown to be incapable of accomplishing his task because it requires the sorting of air molecules into swift and slow and, if the air is in a stable or thermodynamic equilibrium state, there are no such molecules. Of course, Maxwell was unaware of this quantal theorem. He thought of the air molecules in classical terms and, in classical mechanics, the only stable equilibrium state is that of zero momentum and lowest potential energy. In contrast, in the unified theory, equilibrium states, in general, and one stable equilibrium state, in particular, exist at any value of the energy of a fixed amount of air confined in a fixed volume container.

The limitations emerging from the two relatively recent theoretical developments are, of course, facets of the same thermodynamic jewel. For present purposes, what is most noteworthy is that each depends solely on characteristics of a stable or thermodynamic equilibrium state, and is independent of any limitations inherent to the procedures and equipment available to the demon. This observation is universally true. No ultimate limit derived in thermodynamics depends on the specifics of the procedures and equipment used to approach it. To be sure, limitations inherent to any particular practical realization of a demon must be taken into account in addition to the ultimate limit imposed by the characteristic features of a system. But the latter are not as definitive and as decisive as the former.

_________________________________________________________________

*In the case of statistical classical mechanics, the possibility of applying statistical methods to a system having a small number of degrees of freedom was recognized by Gibbs, as well as by E.B. Wilson [10].

V.     CONCLUDING REMARKS

The conception of the demon was motivated solely by Maxwell's conviction that, at the microscopic level, air molecules in a container obey exclusively the laws of classical mechanics, and need not be assigned statistical measures of ignorance, such as the entropy of statistical mechanics.

Despite this explicitly stated conviction, over the past century and a half, all refutations of Maxwell's demon by hundreds of scientists, including Szilard, von Neumann, Prigogine, Feynman, Brillouin, Landauer, Bennett, and Zurek, are based on some kind of statistical entropic or information-theroetic argument that includes both the air molecules and the demon and not, as specified by Maxwell, solely the air molecules. Therefore, all these refutations address neither the problem posed by Maxwell, nor the strange implication of statistics that the degree of information of an observer can influence the ultimate course of physical phenomena.

The exposition of thermodynamics referenced in this paper reveals the existence of a property–the entropy–which is nonstatistical and valid for any system in any state. Thus, it is possible to provide a thermodynamic exorcism of the demon by addressing directly the problem posed by Maxwell.

The unified quantum theory of mechanics and thermodynamics provides an even more dramatic proof of the impossibility of the demon to violate the laws of thermodynamics because there are no fast and slow molecules to be sorted out. So the classical mechanical motivation that precipitated the conception of the demon is eliminated, and he is left without an assignment.